\documentstyle[12pt]{article}

\def\lesssim{\mathrel{\mathpalette\vereq<}}
\def\gtrsim{\mathrel{\mathpalette\vereq>}}
\makeatletter
\def\vereq#1#2{\lower3pt\vbox{\baselineskip1.5pt \lineskip1.5pt
\ialign{$\m@th#1\hfill##\hfil$\crcr#2\crcr\sim\crcr}}}
\makeatother

\def\drawbox#1#2{\hrule height#2pt
        \hbox{\vrule width#2pt height#1pt \kern#1pt
               \vrule width#2pt}
               \hrule height#2pt}

\def\Fund#1#2{\vcenter{\vbox{\drawbox{#1}{#2}}}}
\def\Asym#1#2{\vcenter{\vbox{\drawbox{#1}{#2}
              \kern-#2pt       % line up boxes
              \drawbox{#1}{#2}}}}

\def\fund{\Fund{6.5}{0.4}}
\def\asym{\Asym{6.5}{0.4}}

\begin{document}

\begin{titlepage}
\begin{center}
\today     \hfill    LBNL-40286\\
~{} \hfill hep-ph/9705271\\

%\vskip .25in
\vskip .1in

{\large \bf A Model of Direct Gauge Mediation}\footnote{This work was
  supported in part by the Director, Office of Energy Research, Office
  of High Energy and Nuclear Physics, Division of High Energy Physics
  of the U.S. Department of Energy under Contract DE-AC03-76SF00098
  and in part by the National Science Foundation under grant
  PHY-90-21139, and also by Alfred P. Sloan Foundation.}

\vskip 0.3in
%\vskip 0.1in

Hitoshi Murayama

\vskip 0.05in

{\em Theoretical Physics Group\\
     Ernest Orlando Lawrence Berkeley National Laboratory\\
     University of California, Berkeley, California 94720}

\vskip 0.05in

and

\vskip 0.05in

{\em Department of Physics\\
     University of California, Berkeley, California 94720}

\end{center}

\vskip .1in

\begin{abstract}
  We present a simple model of gauge mediation (GM) which does not
  have a messenger sector or gauge singlet fields.  The standard model
  gauge groups couple directly to the sector which breaks
  supersymmetry dynamically.  This is the first phenomenologically
  viable example of this type in the literature.  Despite the direct
  coupling, the model can preserve perturbative gauge unification.
  This is achieved by the inverted hierarchy mechanism which generates
  a large scalar expectation value compared to the size of
  supersymmetry breaking.  There is no dangerous negative contribution
  to the squark, slepton masses due to two-loop renormalization group
  equation.  The potentially non-universal supergravity contribution
  to the scalar masses can be suppressed enough to maintain the virtue
  of the gauge mediation.  The model is completely chiral, and one
  does not need to forbid mass terms for the messenger fields by hand.
  Beyond the simplicity of the model, it possesses cosmologically
  desirable features compared to the original models of GM: an
  improved gravitino and string moduli cosmology.  The Polonyi problem
  is back unlike in the original GM models, but is still much less
  serious than in hidden sector models.
\end{abstract}

\end{titlepage}

\newpage

Recently, gauge mediation of supersymmetry breaking \cite{oldpapers}
has attracted interests.  The primary motivation of this scheme is to
guarantee degeneracy between the masses of sfermions belonging to different
generations, thereby solving the supersymmetric flavor-changing problem. 
The pioneering works of Dine, Nelson and collaborators \cite{DN,DNS,DNNS}
demonstrated that such a scheme is phenomenologically viable, and they
presented explicit models which realize the gauge mediation mechanism.
Their models, however, require three relatively decoupled sectors
``insulated'' from each other, namely the standard model sector, the
sector of dynamical supersymmetry breaking (DSB), and the so-called
messenger sector.  It is an important question to ask whether this
structure is inevitable for successful gauge mediation or whether it 
simply provides an existence proof, calling for further simplification.

Some progress has been made in this direction
\cite{PT2,AMM1}.\footnote{See \cite{others} for other attempts to
  simplify the structure of gauge mediation models along different
  lines.}  Poppitz and Trivedi \cite{PT2} demonstrated that one can
couple the DSB sector and the standard model gauge groups directly
without spoiling perturbative gauge unification.  However, their model
suffered from two problems.  One is that the supersymmetry breaking
scale is so high that the supergravity contribution to squark and
slepton masses dominate over the contributions from the gauge
mediation.  Therefore, the degeneracy among sfermions is not an
automatic consequence of the model.  Secondly, there are fields which
are charged under the standard model gauge groups below $10^{5}$~GeV,
whose scalar components have supersymmetry breaking soft scalar masses
of $(\mbox{a few}\times 10^{4}~\mbox{GeV})^{2}$.  They contribute to
the two-loop renormalization group equations (RGE) of squark and
slepton masses, driving them negative at low energies \cite{AMM1,APT}.
The problem with supergravity contributions was surmounted in the
model constructed by Arkani-Hamed, March-Russell and the author
\cite{AMM1}, but the problem with light charged fields with large soft
scalar masses afflicts both classes of models.  Despite the problem,
these works left hope that the DSB sector may not need to be as
``insulated'' as in the original models.  The basic ingredient here is
that the mass of the messengers $M_Q^2$ can be much larger than their
supersymmetry breaking bilinear mass $B_Q M_Q$ due to the dynamics of
the models; then the ratio $(B_Q M_Q)/M_Q$ can be kept around
$10^4$~GeV to generate desired magnitude of gaugino masses while $M_Q$
is close to the unification scale, which makes it easy to maintain the
perturbative gauge unification.

In this letter, we present a simple model of gauge mediation without a
messenger sector or gauge singlet fields.  The standard model gauge
groups are an important part of the dynamics.  The model overcomes the
difficulties mentioned above: it preserves the perturbative gauge
unification while suppressing the supergravity contributions enough to
avoid flavor-changing processes.  There are light multiplets in the
DSB sector which are charged under the SM gauge group but they do not
have large soft masses.  Therefore there is no dangerous negative
contribution to the squark, slepton masses due to two-loop RGE, and
the model is phenomenologically viable.  An aesthetically appealing
feature of the model is that it is completely chiral, and one does not
need to forbid mass terms for the messenger fields by hand unlike in
the earlier models.  Since there is no messenger sector in this model,
the vector-like messengers are produced as a consequence of dynamics
(this is a feature also shared by the models in \cite{PT2,AMM1}).
Furthermore, it possesses cosmologically desirable features compared
to the original models of gauge mediation.

Our model is based on the vector-like SP(N) models by Izawa, Yanagida 
\cite{IY} and by Intriligator, Thomas \cite{IT}.  We take SP(4) model 
with 5 flavors, {\it i.e.}\/ 10 fundamentals, which we denote by 
$Q^{i}$ ($i=1,\cdots,10$).  In addition, there are SP(4) singlet fields 
$S_{ij}$ which couple to the $Q$'s in the superpotential,
\begin{equation}
  W = \lambda S_{ij} Q^i Q^j .
\label{eq:Worig}
\end{equation}
This coupling lifts all flat directions: $Q^i Q^j = 0$, while the 
quantum modified constraint requires ${\rm Pf} (Q^i Q^j) = 
\Lambda^{10}$.  The contradiction between two conditions implies that 
supersymmetry is broken.

\begin{table}
\caption[tab1]{The particle content of our model under
  SP(4)$\times$SU(5)$_L \times$SU(5)$_R$ gauge group and a global
  SU(2) symmetry.  The symbol $\cdot$ refers to
  singlets. }
  \begin{center}
    \leavevmode
    \begin{tabular}{c|ccc|c}
      & SP(4) & SU(5)$_L$ & SU(5)$_R$ & SU(2) \\ \hline
      $Q$ & $\fund$ & $\fund$ & $\cdot$ & $\cdot$ \\
      $\bar{Q}$ & $\fund$ & $\cdot$ & $\overline{\fund}$ & $\cdot$ 
      \\ \hline
      $\Sigma$ & $\cdot$ & $\overline{\fund}$ & $\fund$ & $\cdot$ \\ 
      $S$ & $\cdot$ & $\overline{\asym}$ & $\cdot$ & $\cdot$ \\
      $\bar{S}$ & $\cdot$ & $\cdot$ & $\asym$ & $\cdot$ \\ \hline
      $\phi^a$ & $\cdot$ & $\overline{\fund}$ & $\cdot$ & $\fund$ \\
      $\bar{\phi}^a$ & $\cdot$ & $\cdot$ & $\fund$ & $\fund$
    \end{tabular}
    \label{tab:tab1}
  \end{center}
\end{table}

Into the SU(10) global symmetry of the model, we embed
SU(5)$_{L}\times$SU(5)$_{R}$ gauge group.\footnote{This is the gauge
  group above the GUT-scale.  Below the GUT scale, one of the SU(5)
  groups must be broken down to the SU(3)$\times$SU(2)$\times$U(1)
  gauge group to preserve the observed gauge unification.  We use
  SU(5) language to both SU(5) factors to keep track of the quantum
  numbers easily.} The particle content of the model is shown in
Table~1.  The only difference from the original model is that we need
to add $\phi^a$ and $\bar{\phi}^a$ $(a=1,2)$ to cancel the SU(5)
anomalies.  We assume that both SU(5) are weaker than SP(4) and treat
them perturbatively.\footnote{It would be an interesting excercise to
  see how the description of the dynamics changes when one or both of
  the SU(5) groups are stronger than the SP(4) group.}  The
superpotential is nothing but what is obtained by decomposing the
original one (\ref{eq:Worig}) into SU(5)$_{L}\times$SU(5)$_{R}$
multiplets,
\begin{equation}
  W = \frac{1}{2} g S^{kl} Q_k Q_l 
  + \frac{1}{2} \bar{g} \bar{S}_{\kappa\lambda} 
        \bar{Q}^{\kappa} \bar{Q}^{\lambda}
  + \lambda \Sigma_{\kappa}^{k} \bar{Q}^{\kappa} Q_{k} .
\end{equation}
Latin and Greek letters refer to different SU(5) indices.  This is the
most general renormalizable superpotential consistent with gauge
invariance and global SU(2) symmetry.\footnote{This SU(2) can also be
  promoted to a local one.} Just as in the original model, the quantum
modified moduli space requires (at least some of) the meson operators
$QQ$, $\bar{Q}\bar{Q}$ or $\bar{Q}Q$ not to vanish, while the
superpotential requires all meson operators to vanish.  Therefore
supersymmetry is broken.  Note that this model is completely chiral,
{\it i.e.}\/ none of the fields in the model can have mass terms in
the superpotential.  In particular, the absence of terms in the
superpotential such as $S$, $S^{2}$ or $S^{3}$ {\it etc}\/.  had to be
imposed by hand in the vector-like SP(N) models in order to break
supersymmetry, while they are automatically forbidden by the gauge
invariance of the model here.  There are no gauge singlet fields in
this model which is also aesthetically appealing.

We are primarily interested in the situation where $\Sigma$ acquires a
large expectation value.  Existence of a potential minimum with large
expectation value will be shown later.  Consider the following
configuration,
\begin{equation}
  \Sigma^{k}_{\kappa} = \frac{v}{\sqrt{5}} \delta^{k}_{\kappa} =
  \frac{1}{\sqrt{5}} \left( 
    \begin{array}{ccccc}
      v &&&& \\ & v &&& \\ && v && \\ &&& v & \\ &&&& v
    \end{array} \right) ,
    \label{eq:v}
\end{equation}
with $v \gg \Lambda$.\footnote{Here, $v$ is a chiral superfield, but we refer
to its scalar expectation value by the same symbol.}  One can view
this direction as a gauge invariant polynomial det$\Sigma$.  This
configuration breaks SU(5)$_L \times$SU(5)$_R$ gauge group down to
their diagonal SU(5) subgroup.  We identify the unbroken diagonal
SU(5) as ``our'' gauge group into which the standard model gauge
groups are embedded.  Along this direction (\ref{eq:v}), the $\Sigma$
field has an exactly flat potential at the tree-level.  This can be
seen by the following analysis.  First of all, all $Q$'s become
massive with mass $\lambda v/\sqrt{5}$ and can be integrated out from
the theory.  All components of $\Sigma$ except the $v$ direction are
eaten by the broken SU(5) generators.  Therefore the low energy theory
is a pure SP(4) gauge theory with a singlet chiral superfield $v$
together with fields charged under the unbroken diagonal SU(5) $S$,
$\bar{S}$, $\phi$, $\bar{\phi}$.  The low-energy SP(4) gauge group
develops a gaugino condensate, which results in an effective
superpotential
\begin{equation}
  W_{\it \!eff} = \Lambda^2 (\mbox{Pf} M_{ij})^{1/5},
  \label{eq:Weff}
\end{equation}
where the (anti-symmetric) mass matrix $M_{ij}$ of the quarks $Q^i$ 
are given by the expectation values of $\lambda \Sigma$, $g S$ and 
$\bar{g} \bar{S}$,
\begin{equation}
  M_{ij} = \left( \begin{array}{cc}
      g S & \lambda \Sigma \\
      -\lambda {}^{t}\Sigma & \bar{g} \bar{S}
    \end{array} \right) .
\label{eq:Mij}
\end{equation}  
Along the direction of our interest Eq.~(\ref{eq:v}), the effective
superpotential (\ref{eq:Weff}) is
\begin{equation}
  W_{\it \!eff} = \frac{1}{\sqrt{5}} \lambda \Lambda^2 v
  \label{eq:Weffv}
\end{equation}
and the $v$ derivative does not vanish; hence supersymmetry is broken.  
With a canonical (tree-level) K\"ahler potential for the $v$ 
direction, the potential is exactly flat.  On the other hand, the $v$ 
direction acquires an $F$-component, $F_v = \lambda 
\Lambda^2/\sqrt{5}$.  Therefore, the $Q$ fields have a large mass 
$M_{Q}=\lambda v/\sqrt{5}$ with a bilinear supersymmetry breaking mass 
term $B_{Q} M_{Q}=\lambda F_{v}/\sqrt{5}$, and hence act as messengers 
of supersymmetry breaking.  The ratio 
\begin{equation}
  B_{Q} = \frac{F_{v}}{v} = \lambda \frac{\Lambda^2}{\sqrt{5}\, v} 
  \label{eq:BQ}
\end{equation}
determines the size of sfermion, gaugino masses induced by standard
model gauge interactions from the loops of $Q$'s (gauge mediation).
The generated gaugino and scalar masses is $N_{m}=8$ times larger than
the case of the minimal messenger sector ${\bf 5}+{\bf 5}^{*}$ in
\cite{DN,DNS}.  We would like this ratio to be around $B_Q \sim
10^4$~GeV to generate supersymmetry breaking terms of the
desired magnitude.  Recall that $Q$, $\bar{Q}$ are chiral under the
original gauge group; they become vector-like messengers only after
the symmetry breaking SU(5)$_L\times$SU(5)$_R \rightarrow$SU(5).
Since the vector-like messengers originate dynamically in this model,
there is no need for a separate and ``insulated'' messenger sector.

The rest of the mass spectrum of the model is the following at the
renormalizable level (we will discuss additional mass terms generated
by Planck-scale operators later).  The $S$ and $\bar{S}$ fields become
vector-like (${\bf 10}+{\bf 10}^*$) under the unbroken diagonal SU(5)
and acquire an invariant mass.  By expanding the effective
superpotential Eq.~(\ref{eq:Weff},\ref{eq:Mij}) up to the second order
in $S$, $\bar{S}$, one finds that they acquire a mass of $g\bar{g}
\Lambda^{2}/(\sqrt{5}\,\lambda v) = (g\bar{g}/\lambda^{2}) B_{Q} \sim
10^4$~GeV.  $\phi^a$ and $\bar{\phi}^a$ remain massless at this point.
The heavy gauge multiplet has a mass of order $v$, and contributes to
the gaugino masses as well \cite{HGM}.  Two-loop contribution of heavy
gauge multiplet to the scalar masses has not been calculated to the
best of our knowledge.

So far $v$ acquired the $F$-component, but its $A$-component is 
undetermined because its potential is exactly flat at the tree-level.  
In fact, its potential is lifted by the inverted hierarchy mechanism 
\cite{Witten} as follows.  
The K\"ahler potential for the $v$ direction can be calculated by
perturbation theory since it does not participate in strong SP(4)
dynamics, as long as $v \gg \Lambda$.  Its wave function receives
renormalization due to the superpotential coupling $\lambda$ and the
gauge couplings of SU(5)$_L \times$SU(5)$_R$.\footnote{If one follows
  the argument in \cite{Shirman}, one would conclude that there is
  also a contribution from SP(4) gauge coupling to the effective
  potential $V_{\it eff}$.  This is not true because the wave function
  of $\Sigma$ receives renormalization from SP(4) interaction only at
  the two-loop level.  Since the effective superpotential
  Eq.~(\ref{eq:Weffv}) is exact, there is no other effect which
  modifies the effective potential other than the wave function
  renormalization $Z(v)$.  The fact that there is no dependence on
  SP(4) gauge coupling or on the wave function renormalization of $Q$, 
  $\bar{Q}$ is also consistent with the correct treatment
  of Wilsonian renormalization group \cite{AM}.}  From the effective
Lagrangian for the $v$ field
\begin{equation}
        {\cal L} = \int d^{4} \theta Z(v) v^{*} v 
        + \int d^{2}\theta \lambda \Lambda^2 v/\sqrt{5} ,
\end{equation}
the effective potential for $v$ is obtained as
\begin{equation}
  V_{\it eff} = \frac{1}{5 Z(v)} |\lambda \Lambda^2|^2 .
\end{equation}
The point is that the Yukawa contribution and the gauge 
contribution in $Z(v)$ have the opposite sign:  
\begin{equation}
  Z(v) = 1 + 2 \frac{12}{5} \frac{g_L^2+g_R^2}{16\pi^2} \ln \frac{v^2}{M^2}
  - 40 \frac{(\lambda/\sqrt{5})^2}{16\pi^2} \ln \frac{v^2}{M^2}
\end{equation}
Here, $g_L$ ($g_R$) is the gauge coupling constant of SU(5)$_L$
(SU(5)$_R$) group and $M$ the ultraviolet cutoff.  This is the
situation which makes the inverted hierarchy \cite{Witten} possible.
The gauge couplings make $Z(v)$ larger at higher energies which makes
the potential decrease, while the Yukawa coupling makes the potential
increase.  It can well happen that the gauge piece dominates at lower
energies which makes the potential to decrease, but at some point the
Yukawa piece wins over the gauge piece due to the renormalization
group running.\footnote{Consider, for instance, the case where all
  standard model fields couple to SU(5)$_{L}$ which is broken to
  SU(3)$\times$SU(2)$\times$U(1) at the GUT-scale and SU(5)$_{R}$
  remains unbroken between $v$ and the GUT-scale.  Then all $g_{L}$'s
  grow slowly as a function of energy but remain perturbative, while
  $g_{R}$ is asymptotically free.  From the low-energy data,
  $\alpha^{-1}_{3}(v) \sim 8$--$16$ depending on the value of $v$ as
  will be determined later.  Therefore the combination $(1/g_{L}^{2} +
  1/g_{R}^{2}) (v)$ is constrained by the low-energy data, while we
  are free to choose $g_{R}^{2} \gtrsim 2 g_{L}^{2}$.  Then $g_{L}^{2} +
  g_{R}^{2}$ decreases for higher energies.  On the other hand, a
  Yukawa coupling can easily increase as a function of energy as known
  to be the case in the top Yukawa coupling.  It is quite natural that
  a turn-over occurs at some energy scale with $\lambda \sim O(1)$.
  For different embeddings of the standard model generations, the
  details can be somewhat different.  We do not commit ourselves to
  one particular embedding in this letter and do not go into more
  quantiative analysis of the running coupling constants.} The
potential is minimized at the energy scale where this turn-over
occurs.  Because the running is logarithmic in scales, the potential
can develop an expontentially large expectation value for $v$ compared
to the size of supersymmetry breaking given by $\Lambda$.

Even though the minimum of the effective potential discussed above is
a well-defined consistent local minimum, we cannot exclude a
possibility that there is another minimum close to the origin.  For
instance, the point $\Sigma = S = \bar{S}=0$ has an enhanced symmetry
(global U(1)$_R$, which is an accidental symmetry of our model) and is
an extremum of the potential.  Whether it is a minimum or a maximum
cannot be answered because it is in a strongly coupled regime and the
theory is not calculable.  We note, however, that the minimum close to
the origin (if any) is so far away from the other minimum $v \gg
\Lambda$ such that the vacuum tunneling is presumably highly
suppressed, and hence we can live on the minimum discussed.

We have shown that our model which is completely chiral generates 
vector-like messenger fields due to the dynamical breakdown of the 
gauge symmetries.  Because of the inverted hierarchy, $v$ can be 
naturally much larger than $\Lambda$, which makes the messenger 
fields very heavy.  Now we study the values of $v$ 
which give us a viable phenomenology.  First requirement is that the 
potentially non-universal supergravity contribution to the squark, 
slepton masses are suppressed relative to the gauge-mediated 
contribution.  The typical size of the supergravity contribution is 
characterized by the gravitino mass:
\begin{equation}
m_{3/2} = |F_{v}|/(\sqrt{3} M_{*}) 
= B_{Q} v/(\sqrt{3} M_{*}).
\end{equation}
Here, $M_{*}= M_{\rm Planck}/\sqrt{8\pi} \simeq 2\times 10^{18}$~GeV 
is the reduced Planck mass.  We require that the gravitino mass is 
less than 10\% of the gluino
mass $M_{3} = N_{m} (\alpha_{s}(m_{Z})/4\pi) B_{Q}$ such that the 
RGE-induced squark mass squareds have degeneracy at the 1\% level.  
We find 
\begin{equation}
  v \lesssim 0.1 N_{m} (\alpha_{s}(m_{Z})/4\pi) (\sqrt{3} M_{*}) 
  \sim 3\times 10^{16}~\mbox{GeV}.  
  \label{eq:vmax}
\end{equation}
It can be as large as the unification 
scale.\footnote{It is an interesting question whether such a 
large $\Sigma$ is related to the breaking of grand unified SU(5) 
group.  A mechanism for triplet-doublet splitting discussed 
\cite{Barr} based on SU(5)$_{L}\times$SU(5)$_{R}$ may be 
possible within this framework.}

\setcounter{footnote}{0}

So far the $\phi^{a}$, $\bar{\phi}^{a}$ fields are massless, which is
phenomenologically unacceptable.  However, they, together with $S$,
$\bar{S}$ fields, can easily made much heavier by non-renormalizable
interactions suppressed by the reduced Planck mass
$M_{*}$.\footnote{Unfortunately, one possible non-renormalizable term
  in the superpotential det$\Sigma/M_*^2$ not forbidden by the gauge
  invariance could screw up the inverted hierarchy mechanism.  We
  simply assume that it is absent.  Absence of an operator consistent
  with all symmetries of the model is not uncommon in string derived
  models.  Yet higher order term $(\mbox{det}\Sigma)^2/M_*^7$ is not
  harmful.}  Adding the following terms to the superpotential,
\begin{eqnarray}
  \Delta W &=& \frac{1}{M_*^2} \frac{1}{(2! 3!)^{2}}
    \epsilon_{klmnr} \epsilon^{\kappa\lambda\mu\nu\rho}
    S^{kl} \bar{S}_{\kappa\lambda} \Sigma^m_\mu \Sigma^n_\nu
    \Sigma^r_\rho \nonumber \\
    &&+ \frac{1}{M_*^{3}} \frac{1}{(4!)^{2}}
    \epsilon_{klmnr} \epsilon^{\kappa\lambda\mu\nu\rho}
    \phi^{ak} \bar{\phi}^{a}_\kappa
    \Sigma^l_\lambda \Sigma^m_\mu \Sigma^n_\nu \Sigma^r_\rho  ,
\end{eqnarray}
one generates masses for $S$, $\bar{S}$, $\phi$, $\bar{\phi}$ of $m_S
= (v/\sqrt{5})^3/M_*^2$ and $m_\phi = (v/\sqrt{5})^4/M_*^3$.  For
$\phi$, $\bar{\phi}$ to be heavier than the experimental constraints
of order 100~GeV,\footnote{Here we assumed that $\phi$, $\bar{\phi}$
  are stable.  The CDF search for a stable quark \cite{CDF} excluded a
  color-triplet quark bound in unit-charge hadron up to 139~GeV.
  Assuming 50-50 probability for $\phi^a$, $\bar{\phi}^a$ to form
  neutral or unit-charge meson-type states and adding $a=1,2$, this
  can be regarded as the lower bound on $m_\phi$.  LEP-172 has
  excluded stable heavy unit-charged particles close to their
  kinematic reach; for instance the cross section limit from DELPHI
  \cite{DELPHI} can be interpreted as a lower bound of 84~GeV for a
  stable lepton doublet.  Due to the RGE scaling, the bound on
  $m_\phi$ at a high scale determined from these experimental bounds
  is substantially lower, around 50~GeV for both quarks and leptons
  in $\phi$, $\bar{\phi}$.}  we find 
\begin{equation}
  v \gtrsim 3\times 10^{14}~\mbox{GeV}.
  \label{eq:vmin}
\end{equation}
If they happen to be this light, they leave charged tracks in the
detector with anomalous $dE/dx$ because they do not have
renormalizable interactions to the standard model particles due to
the global SU(2) invariance.  On the other hand for the largest
possible $v \sim 3\times 10^{16}$~GeV (\ref{eq:vmax}), $m_{S}
\sim 10^{11}$~GeV and $m_{\phi} \sim 10^{9}$~GeV and then they are
well beyond the experimental reach.  In either case, the gauge
coupling constants remain well perturbative up to the unification
scale with these additional mass terms.  Note that these fields also
contribute to the gaugino and scalar masses when they decouple.
Because of the high power in $\Sigma$ field, their soft supersymmetry
breaking bilinear masses are enhanced.  $S$, $\bar{S}$ fields give
effectively $N_{m} = 9$, and $\phi$, $\bar{\phi}$ fields $N_{m} = 8$.
This makes the degeneracy of squarks, sleptons even better.

It is an interesting question to which SU(5) each of the standard
model fields are coupled above the scale $v$.  The easiest option, of
course, is the case where all standard model fields are coupled to one
of the SU(5), while the other SU(5) has only $Q$, $S$, $\Sigma$ and
$\phi$ as its matter content.  More exotic and interesting possibility
is that the Higgs field and the third generation couple to one SU(5)
while the first and second generation to the other SU(5).  For such a
choice, the Yukawa couplings of first and second generation are
naturally suppressed by a ratio $\langle \Sigma \rangle/M$, where $M$
is a scale which generates non-renormalizable interactions.  It would
be amusing to construct a realistic model of fermion masses along this
line, but it is beyond the scope of this letter.

We now briefly discuss various distinctive features in the
phenomenology of our model compared to the original models of gauge
mediation (see \cite{AMM1} for a more detailed discussion of these
issues).  Since the supersymmetry breaking scale is much higher than
in the original models, the logarithmic running of the supersymmetry
breaking parameters is quite sizable, and the superparticle mass
spectrum is significantly different from the original models because
of these large logarithms \cite{Peskin,DTW}, as well as the large
effective number of messengers $N_{m}=8+9+8=25$.  The experiments will
be able to differentiate these predictions.  The cosmological problem
of a light stable gravitino overclosing the Universe \cite{MMY,Andre}
is much less severe in our model simply because of the larger
gravitino decay constant.  The problem with the coherent oscillation
of string moduli fields \cite{Andre} is also less severe because their
masses are heavier and can be diluted away by a late inflation such as
thermal inflation \cite{LS}.  The cosmological problem of a
Polonyi-like field \cite{Coughlan} is worse than in the original GM
models because of the large scale $v$, but is much less serious than
in the hidden sector models.  The $v$ field acts effectively as a
Polonyi field in hidden sector supersymmetry breaking, with a mass of
order $10^{3}$~GeV.\footnote{The $v$ direction is exactly flat at the
  tree-level, while is lifted due to the inverted hierarchy mechanism
  at the one-loop level.  Therefore, the scalar mass along this
  direction is expected to be $m_{v}^{2} \sim
  (10^{4}~\mbox{GeV})^{2}/16\pi^{2}$.}  But unlike the Polonyi fields
in the hidden sector case, they decay before nucleosynthesis, because
of less suppressed coupling to the light fields $1/v$ rather than
$1/M_{*}$, which provides a crucial improvement.  This can dilute a
pre-existing baryon asymmetry by a factor of $\sim 10^{15} (\langle v
\rangle/M_{*})^{3}$, which is rather mild for $v \lesssim
10^{15}$~GeV.  Even for a larger $\langle v \rangle \sim 10^{16}$~GeV,
the entropy production is still much smaller than in the hidden sector
case $10^{15}$.  Affleck--Dine baryogenesis \cite{AD} may be efficient
enough for this purpose \cite{Andre}.

In summary, we have presented a simple model of gauge mediation using 
the SP(4)$\times$SU(5)$_L\times$SU(5)$_R$ gauge group, where the 
standard model gauge groups is embedded into the diagonal subgroup of 
two SU(5)'s.  The DSB sector is coupled directly to the standard model 
and there is no separate messenger sector.  The model is completely 
chiral, but yet provides vector-like messengers as a result of the 
dynamics.  Despite the direct coupling, perturbative unification can 
be maintained thanks to the inverted hierarchy mechanism.  The 
potentially non-universal supergravity contribution to the squark and 
slepton masses is under control.  We briefly discussed several 
cosmologically desirable features of our model: 
not-too-serious Polonyi problem, and less severe cosmological problems 
associated with the light stable gravitino and string moduli fields 
than in the original gauge mediation models.

\section*{Acknowledgements} I express sincere thanks to Nima 
Arkani-Hamed for many useful discussions crucial for this work.  I 
also thank John March-Russell for collaborations at the early stage of 
this work.  This work was supported in part by the U.S. Department of 
Energy under Contracts DE-AC03-76SF00098, in part by the National 
Science Foundation under grant PHY-95-14797, and also by the Alfred P. 
Sloan Foundation.

\end{document}